\documentclass[%
 reprint,
 nofootinbib,
 amsmath,amssymb,
 aps,
]{revtex4-2}

\usepackage{float}

\usepackage{graphicx}
\usepackage{dcolumn}
\usepackage{bm}

\usepackage{color}

\usepackage{physics}

\graphicspath{{Figures/}}

\begin{document}

\preprint{APS/123-QED}

\title{Black hole induced spins from hyperbolic encounters in dense clusters}

\author{Santiago Jaraba}
 \email{santiago.jaraba@uam.es}
\author{Juan García-Bellido}%
 \email{juan.garciabellido@uam.es}
\affiliation{Instituto de Física Teórica UAM/CSIC, Universidad Autónoma de Madrid, 28049 Madrid, Spain
}%

\date{\today}

\begin{abstract}
The black holes that have been detected via gravitational waves (GW) can have either astrophysical or primordial origin. Some GW events show significant spin for one of the components and have been assumed to be astrophysical, since primordial black holes are generated with very low spins. However, it is worth studying if they can increase their spin throughout the evolution of the universe. Possible mechanisms that have already been explored are via multiple black hole mergers and gas accretion. We propose here a new mechanism that can occur in dense clusters of black holes: the spin up of primordial black holes when they are involved in close hyperbolic encounters. We explore this effect numerically with the Einstein Toolkit for different initial conditions, including variable mass ratios. For equal masses, there is a maximum spin that can be induced on the black holes, $\chi = a/m \leq 0.2$. We find however that for large mass ratios one can attain spins up to $\chi \simeq 0.8$, where the highest spin is induced on the most massive black hole. For small induced spins we provide simple analytical expressions that depend on the relative velocity and impact parameter.
\end{abstract}

\keywords{Black holes; gravitational waves}

\maketitle

\section{\label{sec:introduction}Introduction}

The first gravitational wave detection in 2015 by the LIGO-Virgo collaboration~\cite{PhysRevLett.116.061102} has opened up a new window for our understanding of black holes in the Universe. In particular, the GW190521 event detected in the O3 run of both Advanced LIGO~\cite{AdvLIGO} and Advanced Virgo~\cite{Acernese_2014} has attracted a lot of attention, since assuming that it comes from a black hole merger leads to estimated masses of $91~M_\odot$ and $67~M_\odot$ for the progenitor black holes~\cite{PhysRevLett.125.101102}. The probability that at least one of them is in the range $65-120~M_\odot$ is $99.0\%$~\cite{Abbott_2020}, where there should be a gap in the black hole mass distribution due to pair-instability supernovae~\cite{Farmer_2019}. There are several explanations that have been proposed~\cite{Abbott_2020}, such as a hierarchical merger scenario or that a star with an over-sized hydrogen envelope with respect to its helium core could give rise to such massive black holes. Other possibilities include eccentric mergers, high-mass black hole-disk systems~\cite{PhysRevD.103.063037} or that the black holes involved have a primordial origin~\cite{Garcia_Bellido_1996,Clesse:2017bsw,clesse2020gw190425}, which is the scenario that we will focus on in this paper.

The main problem to this explanation is that primordial black holes (PBHs) are initially generated with low spin~\cite{Chiba:2017rvs}. Therefore, it is interesting to study whether there are spin induction mechanisms for these PBHs, which would then provide a satisfactory explanation for these intermediate mass black holes. In~\cite{PhysRevLett.126.051101}, for instance, it is argued that PBHs could explain the GW190521 signal if they accrete efficiently before the reionization epoch.

One of the possibilities for a Schwarzschild black hole to acquire spin is that it interacts with another one in a close hyperbolic encounter. A numerical exploration of this effect was done by Nelson {\it et al.} in Ref.~\cite{Nelson_2019}, proving that the induced spin could be relevant and reach at least $\chi\approx 0.2$ for equal-mass cases.

We will explore this spin induction effect by studying what happens for different masses. In addition, we will study the trends when we increase the impact parameter and eccentricity, and also when we vary the mass ratio. Finally, we will compare them to what we might expect from a simple analytical approach to this effect.

The layout of this work is as follows. In Sec.~\ref{sec:initialconditions}, we describe the setup that was used for the simulations within the Einstein Toolkit. Later, in Sec.~\ref{sec:results}, we provide the numerical results for the different initial conditions that were treated. A special case of the numerical simulations is the one with mass ratio of 0.1, which is described in Sec.~\ref{sec:q01}. Finally, in Sec.~\ref{sec:comparison}, we compare the trends for the induced spins with some analytical estimations derived in the App.~\ref{sec:analyticalestimation}. We conclude with some final remarks in Sec.~\ref{sec:conclusions}.

Throughout this text, we will work in geometrized units, $G=c=1$.

\section{\label{sec:initialconditions}Grid structure and initial conditions}

In order to simulate black hole hyperbolic encounters in full GR, we have made use of the latest (2020) version of the Einstein Toolkit software~\cite{Loffler:2011ay,EinsteinToolkit}. In particular, the Cactus Computational Toolkit~\cite{Goodale:2002a,Cactuscode:web} was used, as well as the adaptive mesh refinement (AMR) grid infrastructure provided by Carpet~\cite{Schnetter:2003rb,CarpetCode:web}. The initial two black hole data was generated with the TwoPunctures module~\cite{Ansorg:2004ds} with optimized spectral interpolation~\cite{paschalidis2013efficient}. The evolution of the BSSN equations was performed with the McLachlan module~\cite{Brown:2008sb, Kranc:web, McLachlan:web}. Finally, the AHFinderDirect module~\cite{Thornburg:2003sf,Thornburg:1995cp} was used to track the centroids and circumferences of the horizons, the black hole spins were measured with the QuasiLocalMeasures module~\cite{Dreyer:2002mx} and the complex Weyl scalar $\Psi_4$ was determined from the WeylScal4 module~\cite{ZILH_O_2013}.

For this section, whenever we talk about mass, we mean bare mass, as opposed to the ADM mass, which is related but also involves the distance between black holes and initial momenta. The total bare mass $M=m_1+m_2$ in all the simulations will be one, as the software demands, which, together with setting $G=c=1$, completely defines the unit system. We will therefore sometimes omit these quantities, generally keeping the total mass in the expression (e.g. the Schwarzschild radius of a black hole of mass $m$ will be $R_S=2m$).

\subsection{Equal mass case}

The parameter file that has been used is based on the one used in Ref.~\cite{Nelson_2019}. It is similar to the examples for binary black holes provided with the Einstein Toolkit, but with some key modifications that allow increasing the initial separation up to $100~GM/c^2$. For equal masses ($m_1=m_2=M/2\equiv m$), this is a separation of 100 Schwarzschild radii.

\begin{figure}[t]
\centering
\includegraphics[width=0.8\linewidth]{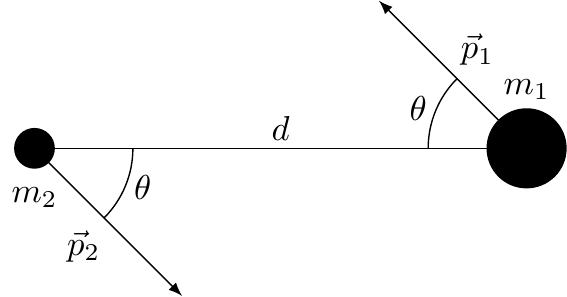}
\caption{\label{fig:initialconditions}Initial conditions in our simulations with Einstein Toolkit, where $m_1\geq m_2$ and $\vec{p}_1=-\vec{p}_2$.}
\end{figure}

For the equal mass case, the initial conditions consist on two black holes located at $x=\pm 50M$, $y=z=0$, with certain initial momenta $\pm\vec{p}$ that in practical terms are controlled via a modulus $p=|\vec{p}|$ and angle $\theta$. The symmetry of the momenta makes sure that the center of mass always lies at the coordinate origin. The situation is depicted in Fig.~\ref{fig:initialconditions}.

The parameters $p/M$ and $\theta$ are related to the usual hyperbolic parameters $b$ (impact parameter) and eccentricity $e$. If the initial momenta were instantaneously translated into initial velocities, we could compute the initial relative speed measured from the rest frame of one of the black holes $V$, as well as the distance $d'$ and angle $\theta'$, where we should account for the Lorentz contraction and time dilation. In this case, one can show that
\begin{equation}
    \label{eq: b from initial conditions}
    b/M=\frac{(d'/M)V}{\sqrt{V^2-\frac{2}{d'/M}}}\sin\theta'.
\end{equation}
\begin{equation}
    \label{eq: e from initial conditions}
    \sqrt{e^2-1}=(d'/M)V\sqrt{V^2-\frac{2}{d'/M}}\sin\theta'.
\end{equation}

However, given that the initial momenta take their time to propagate to the metric quantities and, as a result, to the black hole speeds, we cannot establish such a direct correspondence. Nevertheless, we can interpret an increase in $\theta$ as an increase in both the eccentricity and impact parameter, as the previous equations show.

In order to accommodate these initial conditions, the spatial region is increased to the cube $x,y,z\in[-768M,768M]$. Keeping the resolution while increasing the box size drastically increases the number of divisions, which is not desired. Therefore, it is more reasonable to modify the AMR grid.

Each grid (there is one per black hole) uses half-lengths of $0.75\times 2^n$, for $n=0,1,\ldots,6,8,9,10$. The corresponding steps are $2^n\times\Delta x_{mr}$, now for $n=0,1,\ldots,9$, where $\Delta x_{mr}$ is the size of the most refined grid. Adopting the notation in~\cite{Nelson_2019}, we will refer to $\Delta x_{mr}=(1/56)M$ as ``low'', $(3/200)M\approx (1/66.7)M$ as ``medium'' and $(3/256)M\approx (1/85.3)M$ as ``high'' resolutions.

The time step is initially determined as the spatial step of the bigger grid times a factor \texttt{dtfac} (one of the parameters of the Carpet infrastructure), which we set to 0.05625. Then, this value is divided by a different number on each refinement level, which is controlled via the \texttt{time\_refinement\_factors} array, which we set as [1,~1,~1,~1,~2,~4,~8,~16,~32,~64]. This way, the coarsest four grids are updated at the same rate, and then any finer grid is updated twice as fast as the previous one.

Finally, we can use two symmetries to speed up the code, the most obvious one being the reflection symmetry across the $z$-plane, which is the orbital plane. In addition, for the equal-mass case, the rotating symmetry of $180^\circ$ in the $z$ plane with respect to the origin is also present. Both symmetries reduce the spatial domain by a factor of 4.

\subsection{Changing the mass ratio}

Throughout this article, we will mainly work with $0.7\leq q\leq 1$, where $q=m_2/m_1\leq 1$ ($m_1\geq m_2$). We do this in order to test how the spin induction varies if we do not exactly have the same mass in both black holes, but keeping the ratio close to 1 not to significantly alter the grid structure and the analysis of the problem. The main issues arising for smaller mass ratios will be addressed in Sec.~\ref{sec:q01}.

Generalizing the previous setup for mass ratios of $0.7\leq q\leq 1$ is not very difficult, but there are some things that we have to take into account.

First of all, for a mass ratio of 1, each black hole has half the total mass, which in code units is 0.5. The previous resolutions mean that, per Schwarzschild radius (1 in code units), we have $1/\Delta x_{mr}$ (56, 66.7, 85.3) divisions. However, if we keep the structure for a mass ratio of 0.7, for instance (Schwarzschild radii of 1.18 and 0.82), the number of divisions per Schwarzschild radius is reduced by a factor $1/0.82\approx 1.22$ for the smaller black hole.

Therefore, what we do is adding an extra refinement level for the smallest black hole, in order for its resolution to be better than for the equal mass case. This makes sure that our results are, at least, as good as the equivalent resolution for the $q=1$ case. Also, in order to check that this asymmetry in the extra refinement levels does not introduce errors in the simulation, we have run a few examples with the extra refinement for the $q=1$ case. For the low resolution, the discrepancies between both spins and with respect to the non-refined case are less than $1\%$, which is the typical error involved in simulations with this resolution.

Another thing that is different from the symmetric case is that we must disable the $180^\circ$ rotating symmetry, which essentially doubles the needed computational resources and makes these simulations more time expensive.

Finally, the initial positions are also set to $y=z=0$, with the $x$ so that the initial center of mass is the coordinate origin. We also set $\vec{p}_1=-\vec{p}_2$, as before, to try to keep the center of mass constant. However, due to the mass difference and the fact that the momentum takes some time to stabilize, it is not always satisfied that $m_1\vec{v}_1+m_2\vec{v}_2=0$, which implies that the center of mass $\frac{1}{M}(m_1\vec{r}_1+m_2\vec{r}_2)$ is not completely fixed and moves a bit from the origin. This offset is found to be more relevant for lower values of $q$ and greater values of the initial momentum, as it is reasonable to think. In our case, the center of mass is displaced from the origin, at most, around $5.5M$ during the strong interaction. This does not compromise the final spin measurements, but could have an impact on Weyl scalar-related quantities, such as the gravitational wave strain or the radiated energy. It is, in any case, another reason to be modest with the value of the mass ratio.

\section{\label{sec:results}Numerical results}

First of all, we have run some simulations with equal masses, consisting in different initial incidence angles $\theta$ for the four initial momenta ($p/M=0.245,0.3675,0.49,0.75$ per black hole) considered in Ref.~\cite{Nelson_2019}. For each case, the smallest angle that we consider, $\theta_{\rm min}$, is the one that produces the maximum spin-up according to this article, which is the boundary between hyperbolic and non-hyperbolic events ($e\approx 1$).

In addition, for these four momenta and their corresponding maximum spin-up incidence angles $\theta$, we have run a series of simulations for $0.7\leq q\leq 1$, which is the only parameter that we vary. In particular, we should note that, due to the change in mass while fixing the momenta, the smallest black hole will be faster for $q\neq 1$ than in the equal-mass case, both with respect to the center of mass and to the other black hole.

The dimensionless spin $\chi=a/m$ is computed by using the QuasiLocalMeasures module, which provides the coordinate spin and mass of the black hole. In order to check the consistency of this measurement, we double-check it by comparing to the Christodoulou spin, as it is done in~\cite{Nelson_2019}. We find that both measurements coincide for late times in all cases.

Before showing the results, we will first address their precision.

\subsection{Error analysis}

The differences between the low, medium and high resolutions for the equal mass case were already treated in~\cite{Nelson_2019}. We have double-checked it for some of the highest values of the incidence angle $\theta$, which they do not treat. In particular, the differences between low and medium resolution up to $\theta=4^\circ$ are $<0.5\%$ for $p/M=0.49$, but for $\theta=5.7^\circ$ they rise to $\sim 6\%$. This is probably due to the low induced spin, which begins to be too close to zero ($\sim 0.0004$) and, therefore, the absolute errors involved start to become higher in relative terms. Therefore, the low resolution is enough as long as we take the very low spin values with this caution.

The $q<1$ cases are a bit more complicated for the error analysis. From running simulations of low and medium resolution and both with and without extra refinement level, we can see that the spin measurement of the most massive black hole is very robust ($<2\%$ differences for all cases), but the smallest black hole needs, at least, either medium resolution or the extra refinement level not to present relevant errors (up to $\sim 9\%$ discrepancies). Therefore, we have preferred to generate all the results both with medium resolution and the extra refinement level.

In addition, in order to have an idea of the error of these simulations, we have run the $q=0.7$, $p/M=0.245,0.75$ with high resolution, for which we find maximum differences of order $0.2\%$. This fact, together with the $q=1$ error analysis done in Ref.~\cite{Nelson_2019}, tells us that the differences are smaller than $0.6\%$. Therefore, these will be the typical errors involved in our simulations with varying $q$.

Another thing to mention is that we have also monitored variables that give an idea of whether the simulation is correct or not, such as the Hamiltonian constraint. Due to the enormous storage weight of all the 3D values, we have monitored the average and norms. The results for the Hamiltonian constraint are values of order $10^{-6}$ at most for the 2-norm~\footnote{The $n$-norm is defined by $\norm{A}_n=(\sum |A(i,j,k)|^n/N)^{1/n}$, with $i,j,k$ the spatial grid indices and $N$ the total number of points.}, $10^{-8}$ for the 1-norm and $10^{-10}$ for the average. This is coherent with what is obtained for the cases in~\cite{Nelson_2019} and better than the results for the BBH parameter example of the Einstein Toolkit, which reinforces the idea that the computations are rather accurate. 

\subsection{General behaviour of the simulations}

If we take a look at the time evolution of the spin in any of the simulations, we can see three separate regions. In the first one, we can observe a spin value of approximately zero for both black holes. This is the region where the initial conditions progressively propagate to the metric quantities (the shift is initially zero) as the black hole speeds grow and stabilize, while both black holes progressively get closer.

\begin{figure}[t]
\includegraphics[width=\linewidth]{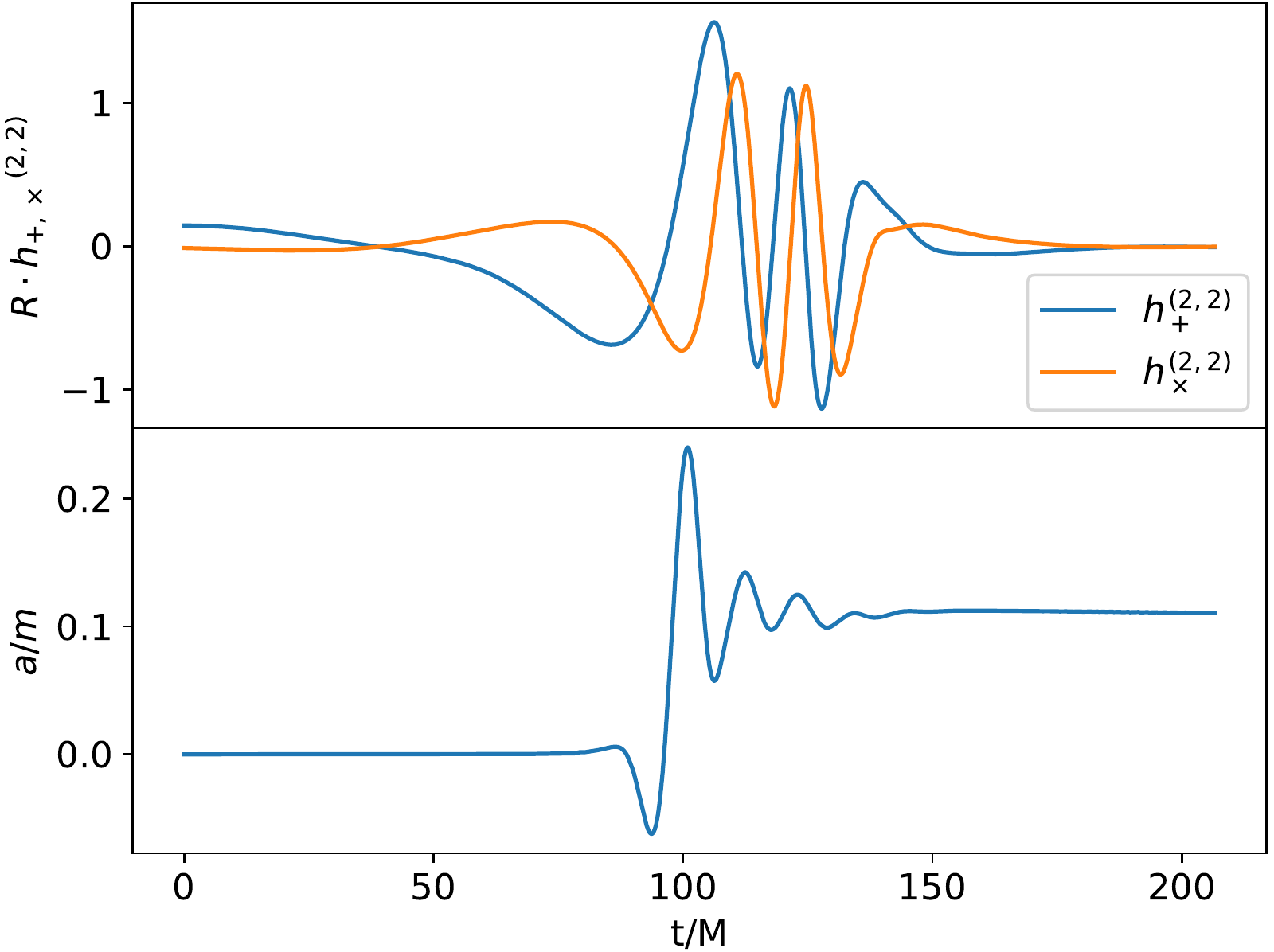}
\caption{\label{fig:strainandspin}Strain of the emitted gravitational wave from the $l=k=2$ multipole (upper panel) and spin evolution (lower) during a hyperbolic encounter with $p/M=0.49$, $q=1$ and $\theta=3.12^\circ$.}
\end{figure}

When both black holes are close enough, they begin to strongly interact and we can see a drastic change on the spins, as well as some oscillations. During this period, some of the energy and angular momentum are radiated away as gravitational waves. We can see that in Fig.~\ref{fig:strainandspin}. The strain amplitude, which will be denoted by $h_{lk}$ ($l\geq 0$, $|k|\leq l$), has been derived from the Weyl scalar on the sphere of radius $R=67.88M$, and has been shifted to compensate for the propagation time to the detector, $\Delta t=-R$. We have used PostCactus~\cite{PostCactus} for this purpose, which computes the strain by using fixed frequency integration and from the expression
\begin{equation}
    h_+^{lk}(r,t)-ih_\times^{lk}(r,t)=\int_{-\infty}^t du\int_{-\infty}^u dv~\Psi_4^{lk}(r,v).
\end{equation}

On the final region, we can see that there is a constant, non-zero spin: the initially non-spinning black holes are now rotating. We measure the final spin at $t=250M$, which is enough for it to have stabilized for all the simulations considered. 

\subsection{Equal masses, varying $\theta$}

First, we treat the four cases $p/M=0.245$, $0.3675$, $0.49$, $0.75$ for $q=1$ and different values of $\theta$, between the maximum spin-up incidence angle and an upper bound $\theta\leq 5.73^\circ$. In order to give an idea of these parameters, we have fit an initial part of the trajectory (from $t/M=30$ to $t/M=80$) to a hyperbola. In Table~\ref{tab:hyperbola}, we provide the ranges of impact parameters $b$ and eccentricities $e$ for the considered cases, as well as the distance of closest approach $r_{\rm min}$. Note that the latter can reach values below $2M$, which would correspond to the sum of the Schwarzschild radii of both black holes, since the apparent horizons of two interacting, rotating black holes are typically smaller, especially when they get close to each other. For these simulations, we get apparent horizon radii of order $R_S/2$ before the strong interaction and $R_S/3$ during it, similarly to what can be seen in other numerical simulations like the one in Fig. 13 in Ref.~\cite{Loffler:2011ay}.

\begin{table}
\caption{\label{tab:hyperbola}Ranges of $\theta$ considered for each initial momentum, as well as the equivalent minimum distances and fitted impact parameters and eccentricities.}
\begin{ruledtabular}
\begin{tabular}{ccccc}
 $p/M$ & $\theta$ (deg) & $r_{\rm min}/M$ & $b/M$ & $e$ \\\hline
 0.245 & 3.47 -- 4.58 & 1.98 -- 4.63 & 6.28 -- 8.30 & 2.15 -- 2.71 \\
 0.3675 & 3.13 -- 4.58 & 1.62 -- 4.90 & 5.72 -- 8.37 & 1.84 -- 2.48 \\
 0.490 & 3.12 -- 5.73 & 1.48 -- 6.79 & 5.78 -- 10.6 & 1.64 -- 2.63 \\
 0.750 & 3.42 -- 5.73 & 1.50 -- 6.36 & 6.61 -- 11.0 & 1.38 -- 1.94 \\
\end{tabular}
\end{ruledtabular}
\end{table}

\begin{figure}[t]
\includegraphics[width=\linewidth]{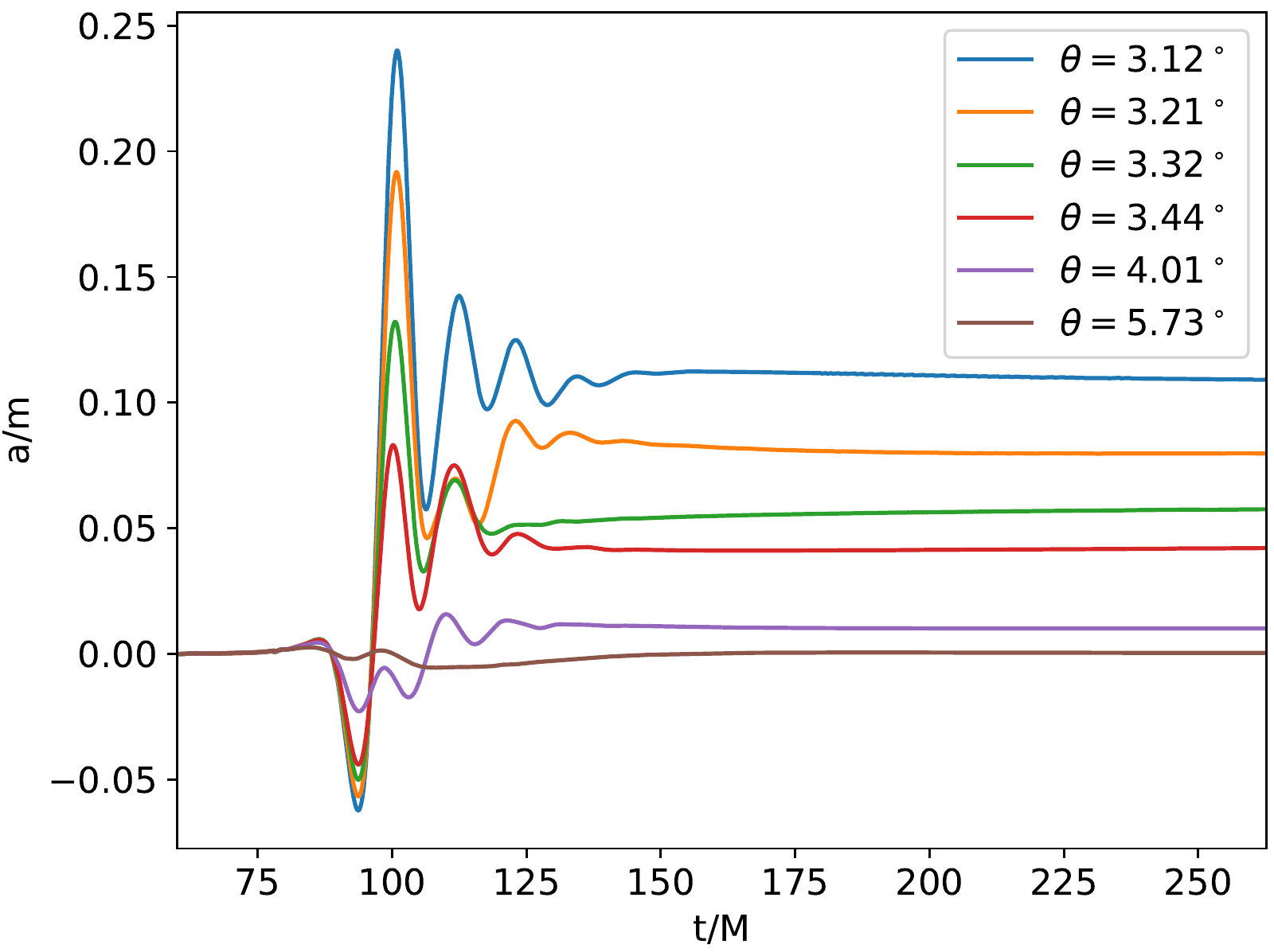}
\caption{\label{fig:spinthetavst}Spin evolution during a hyperbolic encounter with $p/M=0.49$, $q=1$ and different values of $\theta$.}
\end{figure}

Before comparing the final spins in all the cases, we first show the spin evolution versus the time for $p/M=0.49$ in Fig.~\ref{fig:spinthetavst}. We note that the induced spin decreases with growing $\theta$. This is expected, since the closest distance between the black holes increases with the incidence angle, which makes the encounter weaker.

\begin{figure}[t]
\includegraphics[width=\linewidth]{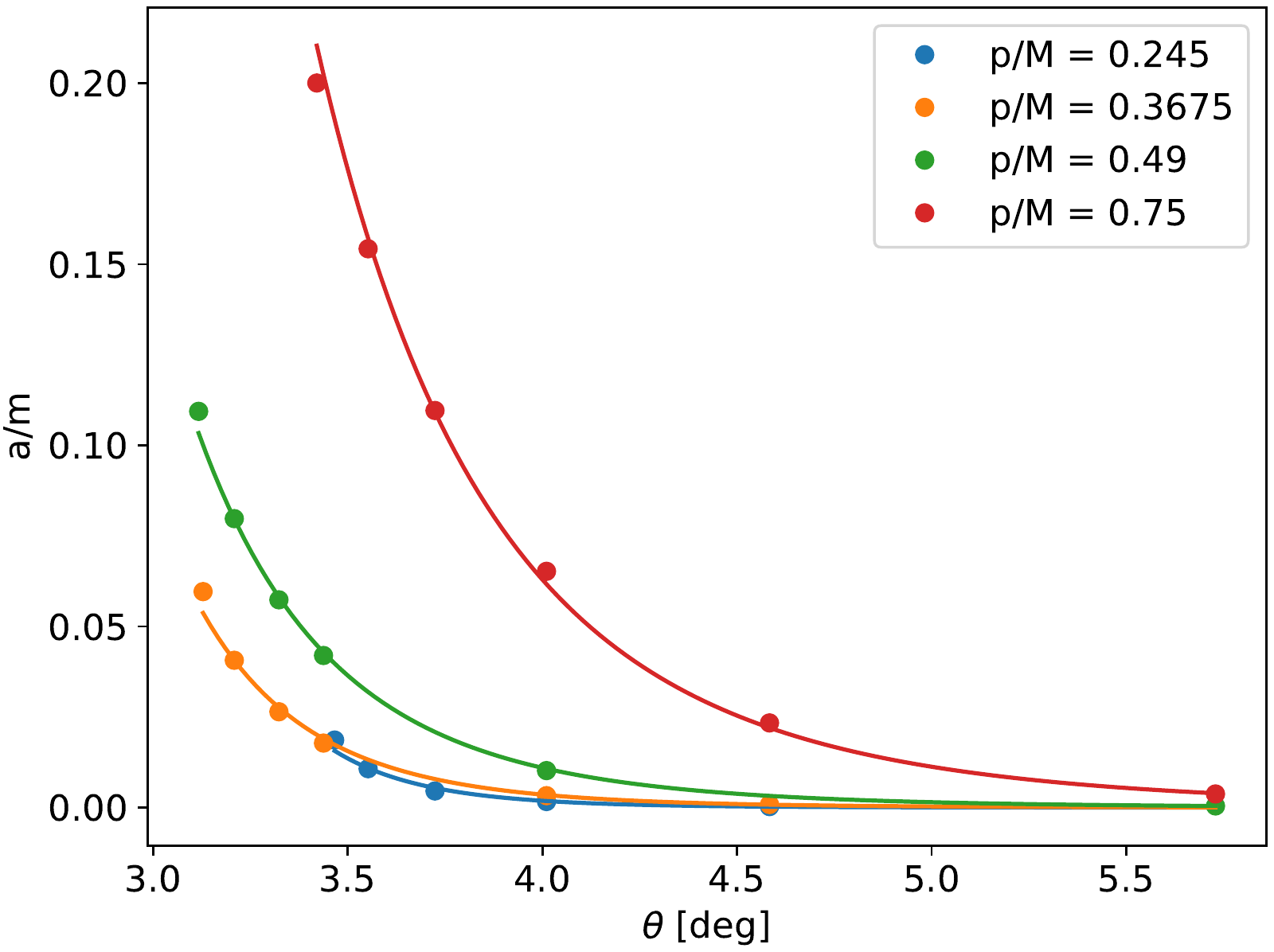}
\caption{\label{fig:spinvstheta}Final spin for hyperbolic encounters with different momenta and $q=1$ versus $\theta$, as well as their fit to a power law.}
\end{figure}

For the four considered initial momenta, we show the final spins versus $\theta$ in Fig.~\ref{fig:spinvstheta}. In particular, one thing we note is that they are reasonably well fitted by a power law. For the $p/M=0.49$ case, the power law also fits well the other points that are shown in Fig. 4a in~\cite{Nelson_2019}, where this spin variation with the incidence angle was already described.

The results of the power law fits are given in Table~\ref{tab:resultstheta}. These were done by linearly fitting the log-log plot, in order to give each point the same importance in terms of relative weight. By doing a least square error fitting to a power law, the smaller values of the spin would have had little impact on the fit.

\begin{table}
\caption{\label{tab:resultstheta}Fitted parameters for Fig.~\ref{fig:spinvstheta} to a power law $\chi=(\theta/\theta_0)^n$, as well as $\theta_{\rm min}$ for reference and the linear correlation coefficient $r^2$ for the ($\log(\theta)$, $\log(\chi)$) data to the corresponding linear function.}
\begin{ruledtabular}
\begin{tabular}{ccccc}
 $p/M$ & n & $\theta_0$ (deg) & $\theta_{\rm min}$ (deg) & $r^2$ \\\hline
 0.245 & -14.8 & 2.62 & 3.47 & 0.9936 \\
 0.3675 & -11.0 & 2.40 & 3.13 & 0.9982 \\
 0.490 & -9.0 & 2.42 & 3.12 & 0.9997 \\
 0.750 & -7.7 & 2.79 & 3.42 & 0.9989 \\
\end{tabular}
\end{ruledtabular}
\end{table}

\begin{figure}[t]
\includegraphics[width=\linewidth]{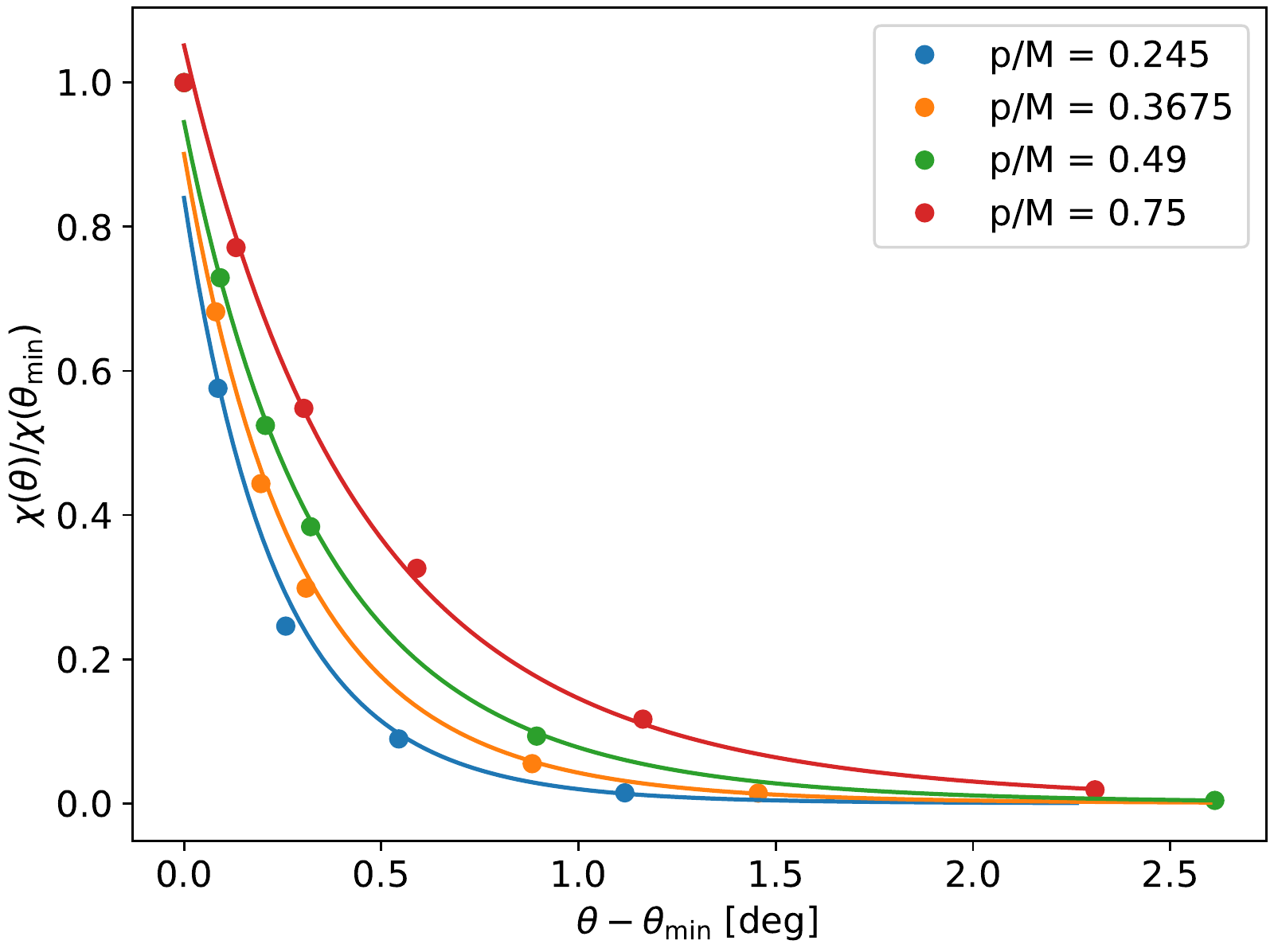}
\caption{\label{fig:spinratiovstheta}Same as Fig.~\ref{fig:spinvstheta}, with the curves normalized by the maximum spin-up value and subtracting $\theta_{\rm min}$ to $\theta$.}
\end{figure}

Finally, in order to better compare these trends, we show the same results in Fig.~\ref{fig:spinratiovstheta}, but now with all the curves normalized by the maximum spin-up value and subtracting $\theta_{\rm min}$ to the incidence angles so that all the curves start from the same point.

\subsection{Varying the mass ratio}

\begin{figure}[t]
\includegraphics[width=\linewidth]{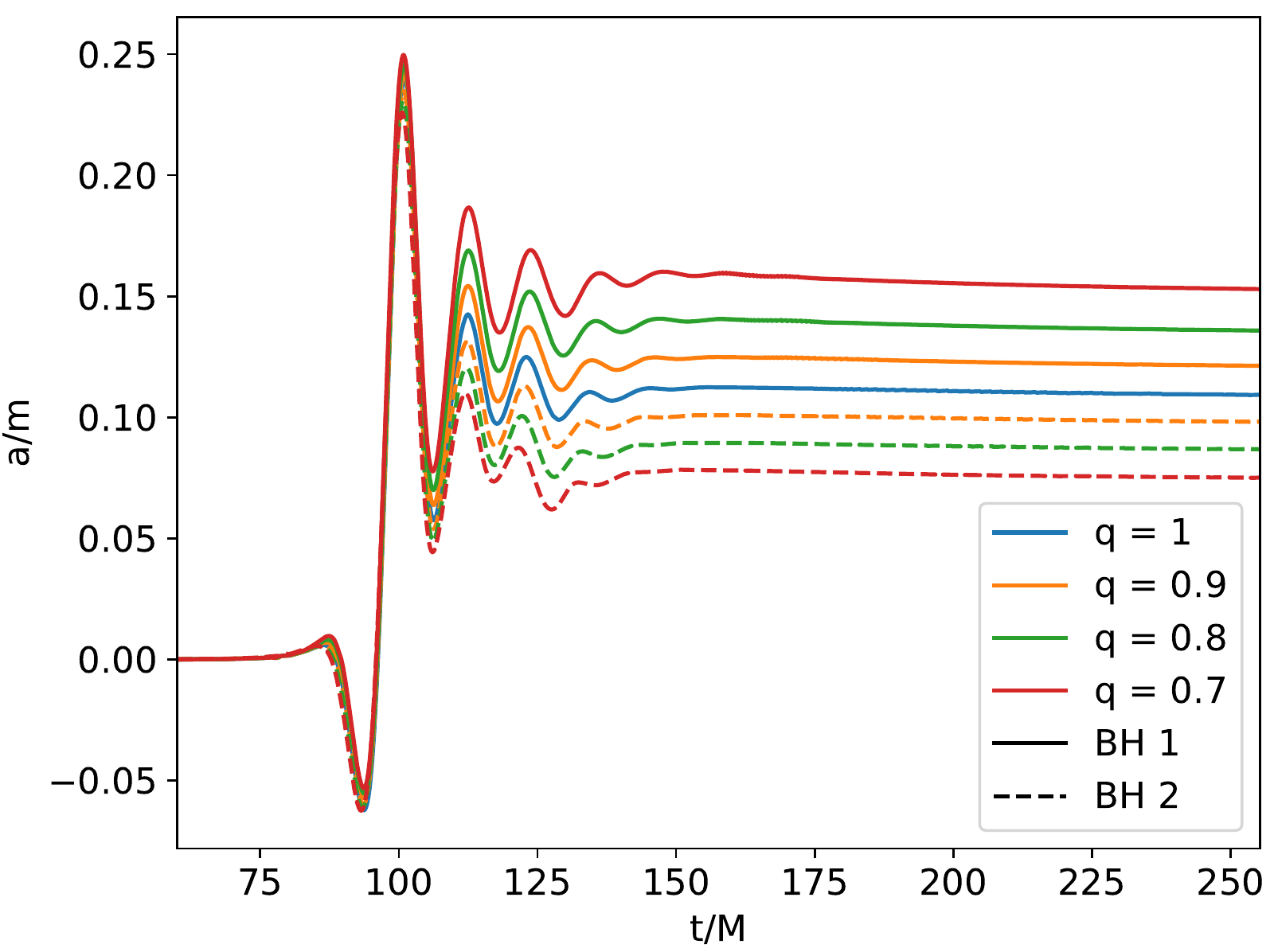}
\caption{\label{fig:spinqvst}Spin evolution during a hyperbolic encounter with $p/M=0.49$, $\theta=3.12^\circ$ and different values of $q$.}
\end{figure}

We will now show the results for the $0.7\leq q<1$ simulations. In Fig.~\ref{fig:spinqvst}, we plot the time evolution of the spin in simulations of different mass ratio, with $p/M=0.49$. Note how each black hole now gets a different spin, where \textit{the highest value is obtained for the most massive black hole}. It is also notorious how the difference between both spins is increased as $q$ gets smaller, as well as the increase of the highest spin and decrease of the smallest one with decreasing $q$.

\begin{figure}[t]
\includegraphics[width=\linewidth]{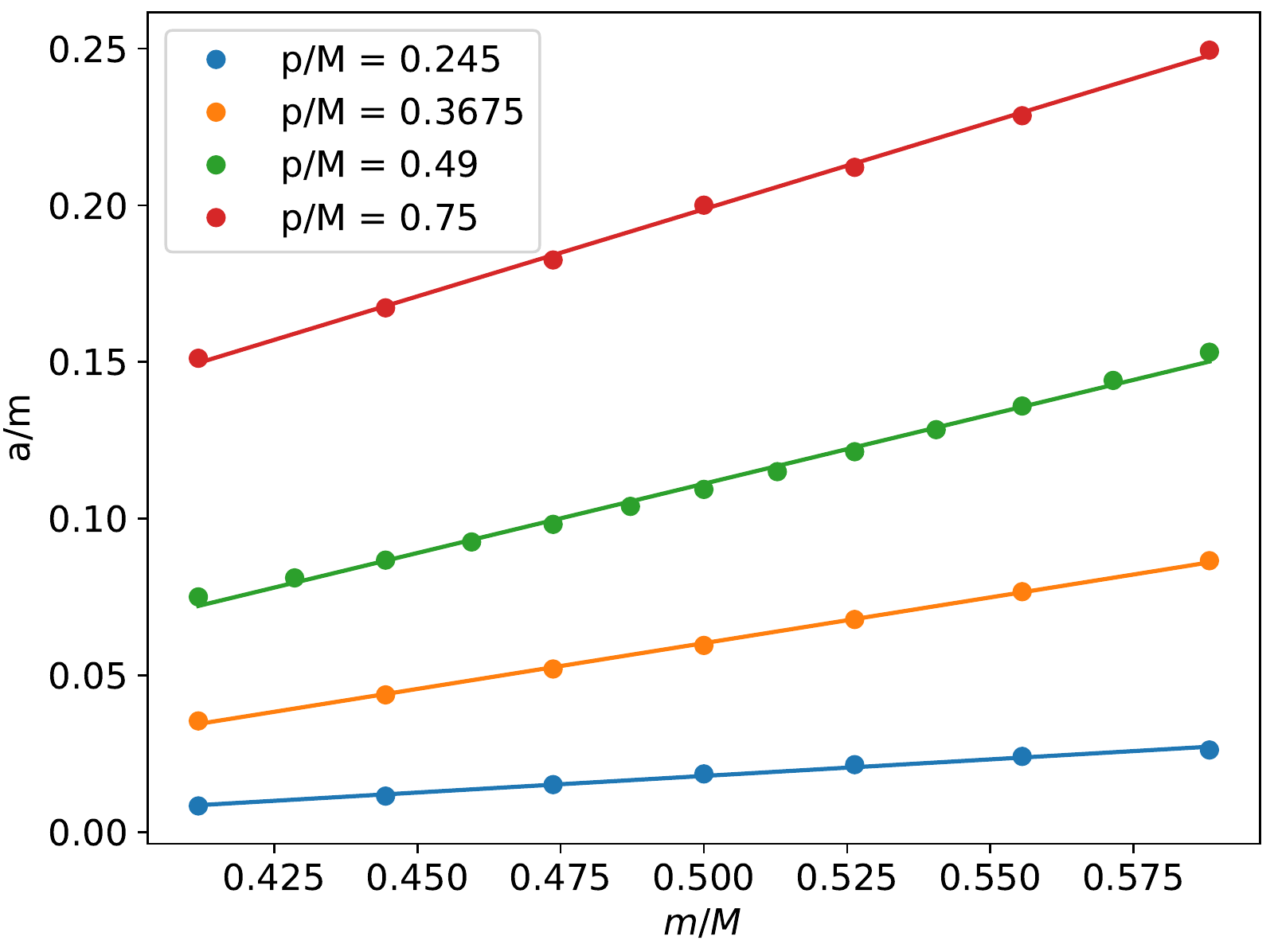}
\caption{\label{fig:spinvsm}Final spin for hyperbolic encounters with different initial momenta and $\theta$ corresponding to the $q=1$ highest spin-up versus the black hole mass, as well as their linear fits.}
\end{figure}

A similar behaviour is observed for the other values of the initial momentum. The final spins are shown in Fig.~\ref{fig:spinvsm}. They are plotted with respect to the masses to avoid having two points per value of the x magnitude, as we would have if we plotted with respect to the mass ratio. Note that the pairs of masses that add up to one come from the same simulation. We can see that they adapt reasonably well to linear fits.

\begin{figure}[t]
\includegraphics[width=\linewidth]{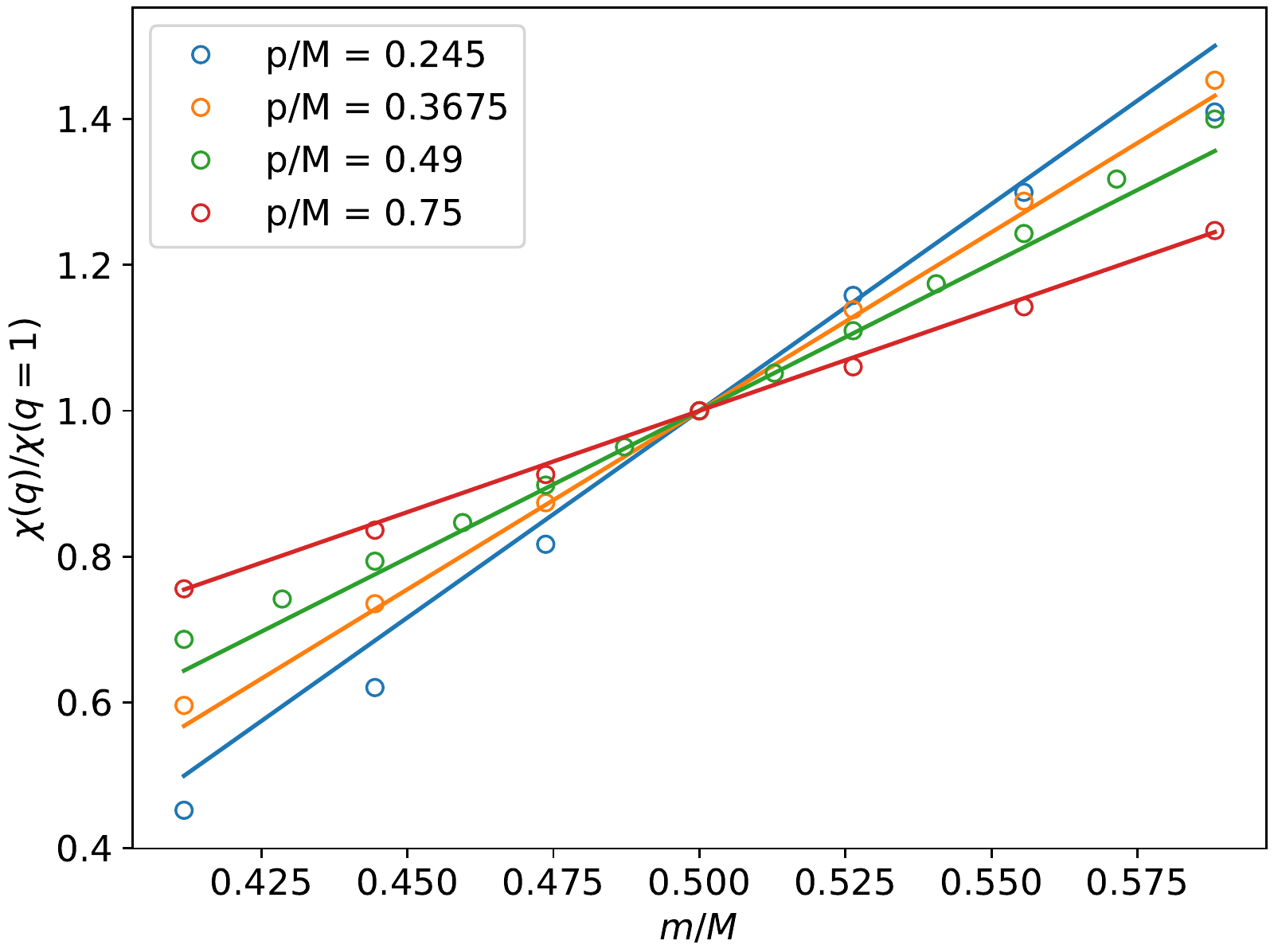}
\caption{\label{fig:spinratiovsm}Same as Fig.~\ref{fig:spinvsm}, where we have divided each value of $\chi$ by the one corresponding to the same initial momentum and $q=1$. The points are now open in order to better see the overlapping values, and the linear fits now have the restriction to pass through the central, common point.}
\end{figure}

In order to better check and visualize how different the trend is for the different initial momenta, we can divide the results by the central value, getting Fig.~\ref{fig:spinratiovsm}. In this case, since the point $m/M=0.5$, $\chi(q)/\chi(q=1)=1$ is common for all the cases, we impose that the linear fits must go through this point and just fit the slope.

We can see that the relative increase between the different values of $q$ is bigger for the smallest values of $p/M$. In addition, the linear fit is generally good, but for the $p/M=0.245$ case is worse than for the other cases. The information about the fits in Fig.~\ref{fig:spinratiovsm} is provided in Table~\ref{tab:resultsq}, as well as the spin of the $q=1$ cases, which can be used to derive the equivalent slope for the $\chi$ vs $m/M$ fit. Here, we can see how the slope decreases with increasing $p/M$.

\begin{table}
\caption{\label{tab:resultsq}Fitted parameters for Fig.~\ref{fig:spinratiovsm}, with their linear correlation coefficient $r^2$ and values of the central spin.}
\begin{ruledtabular}
\begin{tabular}{cccccc}
 $p/M$ & $\theta$ (deg) & slope & $r^2$ & $\chi(q=1)$ \\\hline
 0.245 & 3.46 & 5.7 & 0.979 & 0.0186 \\
 0.3675 & 3.13 & 4.9 & 0.997 & 0.0596 \\
 0.490 & 3.12 & 4.0 & 0.989 & 0.109 \\
 0.750 & 3.42 & 2.8 & 0.997 & 0.200 \\
\end{tabular}
\end{ruledtabular}
\end{table}

\section{\label{sec:q01}Towards lower mass ratios: the case of $q=0.1$}

Finally, we have run a simulation with $q=0.1$, $p/M=0.49$ and $\theta=\theta_{\rm min}\approx 3.12^\circ$. In order to compensate for the loss of (relative) resolution for the smallest black hole, we have added four extra refinement levels to its grid. We have also decided to fix medium resolution. As a result of this configuration, the simulation is much slower than the previous ones.

Unlike the other simulations with $p/M=0.49$ and $\theta=3.12^\circ$, which are hyperbolic, this one ends up producing a merger. One of the possible explanations is that the small black hole starts from a high initial speed, since it has the same momentum as the black holes in other simulations but much smaller mass ($m\approx 0.091$), which would imply more energy loss until its encounter with the heavier black hole. Another possible explanation is a stronger dynamics for $q<1$.

\subsection{Issues with the Weyl scalar}

One of the issues that arise in this simulation is the fact that the center of mass is displaced with respect to the origin. We already mentioned that the maximum deviation for $0.7\leq q<1$ was found to be around $5.5M$, but, in this case, it is around $24M$. This is a problem for the measurements of the Weyl scalar, which are taken at spheres centered at the coordinate origin and, in our case, with radius $r=67.88M$. As a result, computing the strain amplitude of the emitted gravitational wave or its radiated power is also complicated.

This effect has been corrected via a transformation of the Weyl scalar from the sphere centered at the origin ($S_0$) to the sphere centered at $\vec{r}_{\rm CM}(t-R/c)$ ($S_{\rm CM}$) for each time $t$. In order to do this transformation, one has to:
\begin{itemize}
    \item Convert the available $\Psi_4$ multipoles (in our case, up to $l=4$) to a scalar field defined at $S_0$.
    
    \item For each $(t, \vec{p}(t))$, with $\vec{p}(t)\in S_{\rm CM}$, get the light ray that originated at $(t-R/c, \vec{r}_{\rm CM}(t-R/c))$ and passes through $\vec{p}$ and take the value $r\Psi_4$ when it passes through $S_0$. Then, divide by $R$ to get the value of $\Psi_4$ at the desired point.
    
    \item Convert the resulting scalar field at $S_{\rm CM}$ back into multipoles.
\end{itemize}

\begin{figure}[t]
\includegraphics[width=\linewidth]{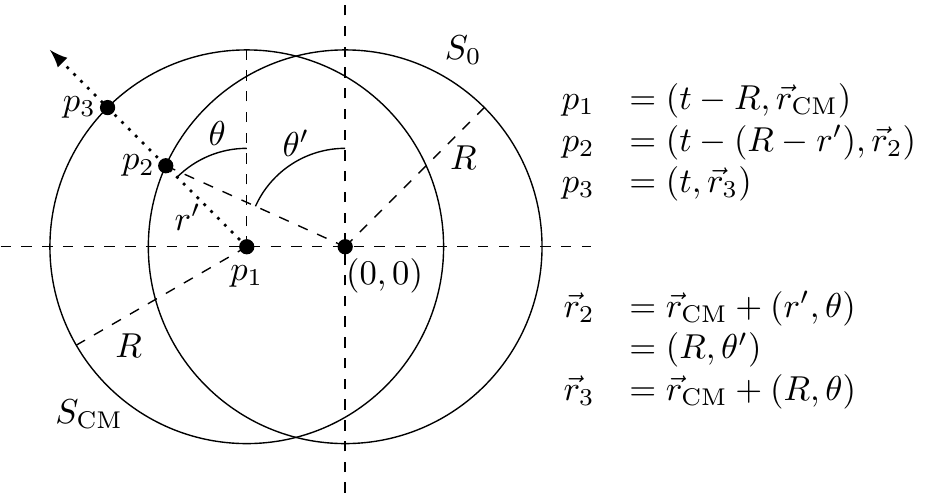}
\caption{\label{fig:spheretransf}All the relevant points and quantities involved in the transformation of the Weyl scalar from the sphere $S_0$ to the sphere $S_{\rm CM}$, for an arbitrary space-time point $p_3=(t,\vec{r}_3)$, $r_3\in S_{\rm CM}(t-R)$. The dotted line represents the light ray which passes through the three relevant points $p_i$, $i=1,2,3$.}
\end{figure}

In practice, we have a grid $(t, \theta, \varphi)$ for $S_{\rm CM}$ and need the equivalent points $(\theta', \varphi')$ at $S_0$, its radius $r'$ with respect to $S_{\rm CM}$ and the time at which the light ray passes through it, $t-(r'-R)/c$. The situation is depicted in Fig.~\ref{fig:spheretransf}.

Note, however, that this correction is far from perfect. First, this assumes that a light ray exactly propagates through the coordinates at speed $c=1$, while the space-time curvature can slow down this speed. In addition, we have only produced up to the $l=4$ multipole, which is enough for a sphere centered at the origin but, in this case, the contribution of the multipoles $l\geq 5$ measured at $S_0$ could be non-negligible even for multipoles $l\leq 4$ at $S_{\rm CM}$.

\begin{figure}[t]
\includegraphics[width=\linewidth]{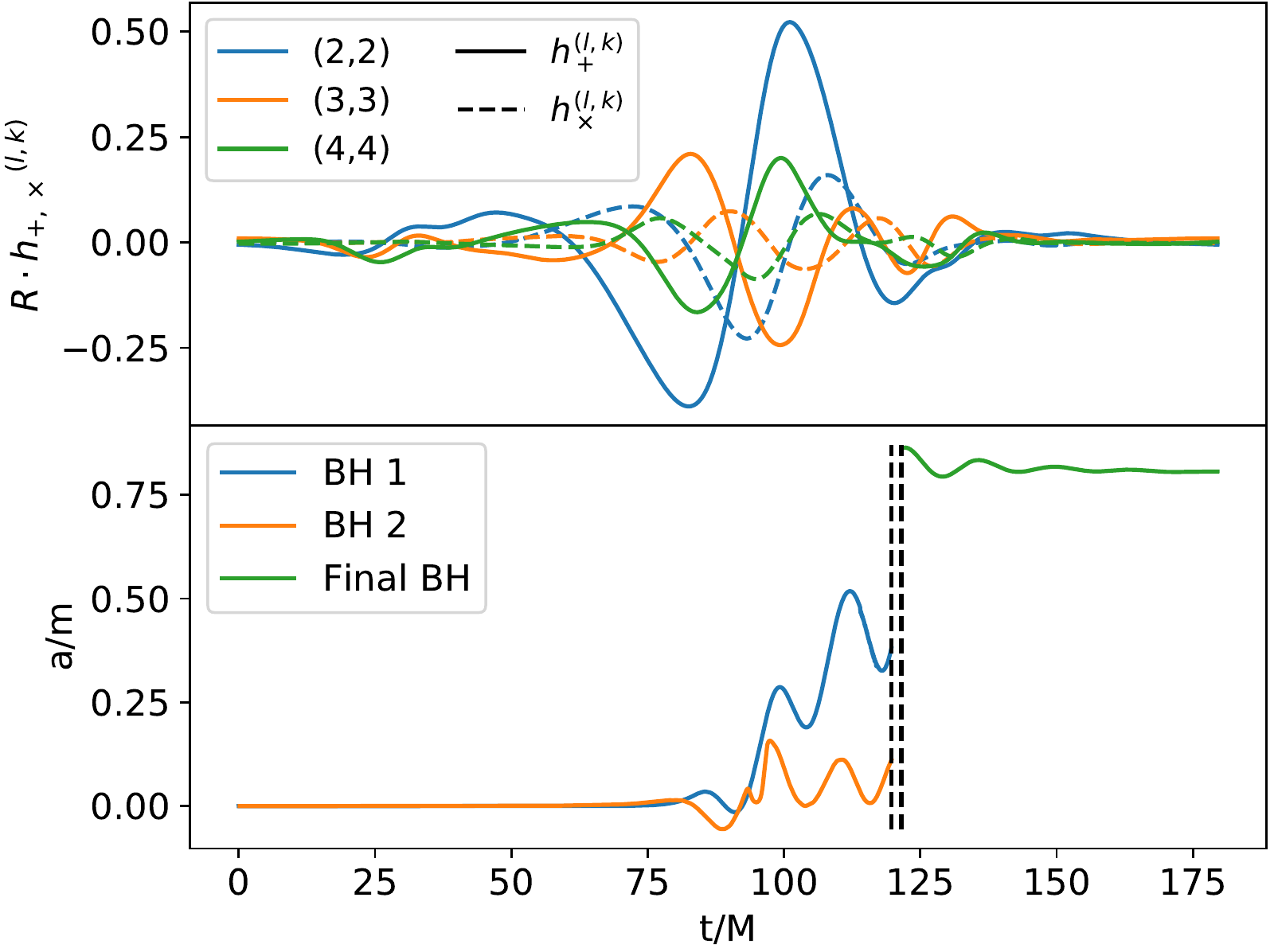}
\caption{\label{fig:strainspinq01}Strain amplitudes of the $q=0.1$ simulation for the modes $l=k\leq 4$ (upper panel), together with the spin evolution (lower), where the first black hole is the most massive one. The dashed, vertical lines separate the periods before and after the merger.}
\end{figure}

Correctly measuring the Weyl scalar is important to determine some gravitational wave-related quantities, such as its strain amplitude. We show the amplitudes for the modes $l=k\leq 4$ in Fig.~\ref{fig:strainspinq01}, with the corresponding time shift of $-R$, together with the spin evolution.

\subsection{The spins}

Even if this simulation ends up in a merger, we can observe the spin evolutions and draw some conclusions. First, we can see how the spin-induction phenomenon works in the same way as in other cases: higher spin is induced on the heaviest black hole. This can be observed in the lower panel of Fig.~\ref{fig:strainspinq01}, which represents the temporal spin evolution.

In this figure, we have drawn two dashed, vertical lines. The first of them marks the moment when the centroids of both apparent horizons are at a distance equal to the sum of their mean radii. This means that both black holes are already too close to continue tracking each of them separately. The second vertical line marks the moment from which the joint horizon can be followed.

One of the main conclusions that should be drawn from this simulation is that two initially non-spinning black holes involved in a hyperbolic encounter that ends in merger can naturally acquire a relevant spin while they approach each other. In practice, this means that, if we observe only the last oscillations of a merger through its emitted gravitational waves, and estimate their initial spins from an inspiral waveform template, \textit{we cannot assume that they had this large spin asymptotically away from the merger}. If they started as a hyperbolic event, they could have acquired their spin as they scattered off each other, emitting gravitational waves and becoming a bounded system that finally merged in a few oscillations.

\section{\label{sec:comparison}Comparison with analytic expressions}

The problem of analytically deriving the induced spins in a close hyperbolic encounter is not trivial. In fact, in order to accurately predict the results of our simulations, one would have to get to, at least, PPN(4) order~\cite{PhysRevD.51.5753}. This strong field interaction cannot definitely be modelled with weak field approximations, as the minimum separations of order $1$--$7M$ in Table~\ref{tab:hyperbola} show.

Nevertheless, one can take some naive approaches to this question in order to, at least, see whether we can qualitatively predict the trends or not. This is what we try to do in this section with the two expressions derived in the App.~\ref{sec:analyticalestimation}, namely~\eqref{eq: chi ring} and~\eqref{eq: chi PN 1}, \eqref{eq: chi PN 2}.

\subsection{Trend for varying $\theta$}

First of all, we will study what happens with $\theta$. If we convert the spin expressions as functions of $v_\infty=v_0\sqrt{(e-1)/(e+1)}$, which can be assumed to be common for equal values of $p/M$ and $q=1$, we get
\begin{equation}
    \chi=f\times\frac{(e+1)^{1/2}}{(e-1)^{5/2}}v_\infty^5,
\end{equation}
where $v_\infty$ is the asymptotic velocity between both black holes at infinity and with $f$ given by
\begin{equation}
    \label{eq: f ring}
	f_{\rm ring}=\frac{16}{5}\eta
\end{equation}
for the ring expression and by
\begin{equation}
    \label{eq: f PN 1}
    f_{\rm 1.5PN, 1}=\left[\frac{4}{5}\eta+\frac{6}{5}(1-\sqrt{1-4\eta})\right]\frac{m_1}{M},
\end{equation}
\begin{equation}
    \label{eq: f PN 2}
    f_{\rm 1.5PN, 2}=\left[\frac{4}{5}\eta+\frac{6}{5}(1+\sqrt{1-4\eta})\right]\frac{m_2}{M}
\end{equation}
for the PPN(1.5) approximation.

In order to relate the previous expressions with $\theta$, we can use the expression~\eqref{eq: e from initial conditions}. By neglecting the Lorentz contraction and since the constants involved are the same for all the cases with fixed initial momentum, we conclude that $\sin\theta\propto\sqrt{e^2-1}$. Therefore, for low values of $\theta$, we can use $\theta\propto\sqrt{e^2-1}$.

By neglecting the missing factors, we can assume that $\theta\ll 1$ implies $e\gtrsim 1$, which implies treating the $e+1$ factors as constants. By doing this, we find that our equations can explain a trend $\chi\sim\theta^{-5}$. This is not enough to predict the exponents that we observe in the results (Table~\ref{tab:resultstheta}), but the difference could be easily explained as the missing strong field interaction that we do not take into account, which underestimates the spin for lower $\theta$ (lower impact parameter and eccentricity).

\subsection{Trend for varying $q$}

According to the ring expression~\eqref{eq: chi ring}, we would not expect different final spins for both black holes of a given simulation. However, the PPN(1.5) expressions~\eqref{eq: chi PN 1},~\eqref{eq: chi PN 2} do predict different spins. In fact, they accurately predict the fact that the greater spin is induced in the most massive black hole.

Another success of the PPN(1.5) approximation is that it predicts that the induced spin on a given black hole is directly proportional to its mass. This is what we see in the results, particularly in Figs.~\ref{fig:spinvsm} and~\ref{fig:spinratiovsm}. By dividing by the central spin, as in this second figure, and assuming that the relative speed is the same in all the cases, we can cancel out some constants and get a straight line with slopes
\begin{equation}
    \frac{4}{7}\left[2\eta+3\left(1-\sqrt{1-4\eta}\right)\right]m_1,
\end{equation}
\begin{equation}
    \frac{4}{7}\left[2\eta+3\left(1+\sqrt{1-4\eta}\right)\right]m_2,
\end{equation}
for the most and least massive black holes, respectively. For $q\sim 1$, both slopes are 1, which at least reproduces the order of magnitude of the slopes in Table~\ref{tab:resultsq}, but the difference is clear. Again, this can be explained because our approximation does not tackle strong-field interactions.

Also, this time, there is another source of error, which is that the numerical constants that we pretend to get rid of by dividing by the central spin do not disappear. Since they involve speeds, they are different for each value of $q$, which could impact the trends.

\section{\label{sec:conclusions}Conclusions}

Following the work of Ref.~\cite{Nelson_2019}, we have shown that it is possible to induce spins in two initially non-spinning, equal-mass black holes. They are larger for higher initial velocities and smaller values of the impact parameter.

In addition, we have studied hyperbolic encounters where the two black holes have different masses and
found that, for a given impact parameter and initial relative velocity, the highest spin is induced on the most massive black hole. In particular, we find that the spin induction effect can be significantly enhanced for the most massive black hole when the mass ratio becomes large. This new result suggests a viable mechanism for significant spin induction in PBHs, contrary to the case of gas accretion where the induced spins cannot acquire large values.
    
Furthermore, we are able to qualitatively predict the trends of the spin with varying impact angle and mass-ratio with simple weak-field approximations. However, in order to get more accurate predictions of the induced spins, one would have to resort to higher orders of the PPN formalism. This is left for future work.
    
With our expressions for the induced spins, we might expect more accurate predictions for more modest values of the involved parameters (weaker interaction). However, they are difficult to generate with the Einstein Toolkit, since the errors involved in low spin measurements are higher in relative terms. Also, the interaction times could get significantly bigger and we would need larger separations, which might be problematic from a technical point of view. Nevertheless, with enough computing power and time, these simulations are possible and should be explored in the future.

Finally, we have found that two initially non-spinning black holes involved in a hyperbolic encounter with intermediate mass ratio ($q\sim 0.1$) that ends in a merger, can naturally acquire a relevant spin, $\chi\simeq 0.8$, for the more massive black hole. This result is relevant for the interpretation of some of the events like GW190521 found by LIGO/Virgo, since the progenitors could have started being very massive but spinless primordial black holes.

We note that most of the hyperbolic encounters in dense PBH clusters occur at large impact parameters (many times their Schwarzschild radius) and small relative velocities ($v_0 \ll c$), and therefore the induced spin will be negligible for the majority of the black holes in the cluster. However, from time to time, a hyperbolic encounter between a large-mass-ratio pair will spin-up the more massive PBH to values of $\chi$ significantly different from zero, up to $\chi\leq0.2$. This could explain why we observe in LIGO/Virgo GW events~\cite{Garcia-Bellido:2020pwq} a distribution of spins peaked around zero with dispersion $\Delta\chi \sim 0.2$. A more refined study taking into account the distribution of eccentricities, impact parameters and relative velocities in dense PBH clusters~\cite{trashorras2020clustering} should give us a prediction for the expected spin distribution depending on the mass and compactness of the cluster. We leave this for a second publication.

\begin{acknowledgments}
We would like to thank Zachariah Etienne and Patrick E. Nelson for their help regarding BH hyperbolic encounters with the Einstein Toolkit. All the simulations have been run in the Hydra HPC cluster in the IFT. S.J. is supported by a predoctoral contract by the Spanish Ministry of Science, Ref.~SEV-2016-0597-19-2. The authors acknowledge support from the Research Project PGC2018-094773-B-C32 [MINECO-FEDER], and the Centro de Excelencia Severo Ochoa Program SEV-2016-0597. 
\end{acknowledgments}

\appendix

\section{\label{sec:analyticalestimation}Analytic estimate of the induced spins}

Estimating the spin induced on the black holes participating in a close hyperbolic encounter is a complicated issue. In order to get an accurate analytical estimation, one would have to get to, at least, PPN(4) in the post-Newtonian formalism.

However, we can approach this problem by addressing the frame dragging involved in this dynamics. The precession vectors can be interpreted as the angular speeds that are induced on the corresponding inertial frames. Therefore, they will be our best guess to estimate induced spins without resorting to higher orders in the PPN formalism.

\subsection{Ring approximation}

From the rest frame of a black hole, a close encounter with another one with a certain mass $m$ is just a point mass current following a certain trajectory $\vec{r}(t)$, which is exactly a hyperbola in the keplerian limit. This situation is analogous to a black hole located at the center of a massive ring of mass $m$, with a time-varying radius following the equation $R(t)=|\vec{r}(t)|$ and rotating so that the speed of each of its points matches the velocity that the point mass would have.

To simplify things, we will assume that the induced spin on the central black hole has the same order of magnitude if we take the stationary situation where $R$ and $\omega$ are constant, with their values corresponding to the point of closest approach.

First of all, we consider a thin ring of certain mass $m_1$ and radius $R$, which rotates around its axis at a certain angular speed $\omega$. By going through the PPN(1.5) formalism, one can show that the central black hole undergoes a certain precession given by the vector
\begin{equation}
    \vec{\Omega}=\frac{2}{r^3}\vec{J},
\end{equation}
where $\vec{J}$ is the angular momentum of the ring. It is interesting to note that this expression exactly matches the one for the precession of the orbital angular momentum of a test particle orbiting a rotating black hole, an effect which is known as Lense-Thirring precession~\cite{Lense:1918zz}.

For a thin ring, the moment of inertia with respect to the center is just $I=m_1R^2$. Therefore,
\begin{equation}
    \vec{\Omega}=\frac{2}{r^3}\vec{J}=\frac{2}{R^3}m_1R^2\vec{\omega}=\frac{2m_1}{R}\vec{\omega},
\end{equation}
which relates the angular speed of the ring with that of the induced inertial frame at its center. The same relation between $\vec{\Omega}$ and $\vec{\omega}$ can be obtained from the equations in~\cite{Horatschek_2010} in the thin-ring approximation.

We can now compute the dimensionless spin that would correspond to a black hole of certain mass $m_2$ located at the center of the ring. For this purpose, we will assume that the frame-dragging speed $\Omega$ is completely transferred to the central black hole, which rotates with this angular speed. If we take the black hole to be a solid sphere whose radius coincides with its Schwarzschild radius $R_{S,2}$, its moment of inertia would be
\begin{equation}
	I_2=\frac{2}{5}m_2R_{S,2}^2=\frac{2}{5}m_2(2m_2)^2=\frac{8}{5}m_2^3.
\end{equation}

With this, we can get the dimensionless spin,
\begin{equation}
	\chi=\frac{a}{m_2}=\frac{J}{m_2^2}=\frac{I_2\Omega}{m_2^2}=\frac{16}{5}\frac{m_1m_2}{R}\omega.
\end{equation}

Finally, we want to extrapolate this result to the hyperbolic motion of a black hole of mass $m_1$ around a black hole of mass $m_2$. For this purpose, we will express $\omega$ and $R$ in terms of two parameters which characterize the hyperbolic motion: the velocity at the point of closest approach, $v_0$, and the eccentricity of the orbit, $e$.

First of all, we take the radius of the ring, $R$, to be the distance between both black holes at the moment of closest approach. In hyperbolic motion, this distance is given by $R=a(e-1)$, where $a$ is the semimajor axis of the hyperbola and must not be confused with the dimensionless spin $a$.

Second, when we substitute the ring by a black hole, we keep the same angular momentum. Therefore,
\begin{equation}
	L_{\rm ring}=L_{\rm BH}\Rightarrow m_2R^2\omega=m_2Rv_0\Rightarrow \omega=v_0/R.
\end{equation}

We can also express the semimajor axis in terms of $v_0$. First, we can write (take e.g.~\cite{Garcia_Bellido_2017, Garcia_Bellido_2018})
\begin{equation}
	bv_\infty^2=M\sqrt{e^2-1},
\end{equation}
where $b$ is the impact parameter ($b=a\sqrt{e^2-1}$) and $M$ the total mass of the system, $M=m_1+m_2$. By conservation of angular momentum,
\begin{align}
	\nonumber
	L_0=L_\infty&\Rightarrow a(e-1)v_0=bv_\infty\Rightarrow v_0=v_\infty\sqrt{\frac{e+1}{e-1}}\\
	\label{eq: a v0}
	&\Rightarrow a^2v_0^4=\frac{b^2}{e^2-1}v_\infty^4\frac{(e+1)^2}{(e-1)^2}=M^2\frac{(e+1)^2}{(e-1)^2}.
\end{align}

Putting all this together, we get
\begin{align}
	\nonumber
	\chi &= \frac{16}{5}\frac{m_1m_2}{R^2}v_0=\frac{16}{5}\frac{m_1m_2}{a^2(e-1)^2}v_0\\
	\label{eq: chi ring}
	&=\frac{16}{5}\frac{m_1m_2}{M^2}\frac{1}{(e+1)^2}v_0^5.
\end{align}

For the particular case $m_1=m_2$, the previous expression is just
\begin{equation}
	\label{eq: equal mass ring}
	\chi=\frac{4}{5}\frac{1}{(e+1)^2}v_0^5.
\end{equation}

For the next case, it is useful to note that, in order to convert from $\Omega$ to $\chi$, we have just multiplied by a factor
\begin{equation}
	\frac{\chi}{\Omega}=\frac{I}{m_2^2}=\frac{8}{5}m_2.
\end{equation}

\subsection{Mass current}

Another possible approach to compute the spin is introducing a mass $m_1$ current at a position $\tilde{\vec{r}}$ with speed $\tilde{\vec{v}}$. We start by writing the angular velocity of an inertial frame within the gravitational potentials $\phi$ and $\vec{g}$, which can be taken from the equation (9.6.12) in Ref.~\cite{Weinberg},
\begin{equation}
	\label{eq: Omega curl}
	\vec{\Omega}=-\frac{1}{2}\nabla\times\vec{g}-\frac{3}{2}\vec{v}\times\nabla\phi.
\end{equation}
The second term corresponds to the de Sitter effect, coming from the gravitoelectric part of the potential. If we assume that the mass current is symmetrically distributed within a ring, then the potential $\phi$ at the center is constant and we can safely ignore this component.

Therefore, we just have to compute $\vec{g}$, which can be done with the expression
\begin{equation}
	\vec{g}=-4G\int d^3\vec{r'}\frac{\rho(\vec{r'})\vec{v}(\vec{r'})}{|\vec{r}-\vec{r'}|}.
\end{equation}

We will now substitute $\rho$ and $\vec{v}$ by the ones corresponding to a point mass current at position $\vec{\tilde{r}}(\varphi)$, where $\varphi$ is an angular variable that parametrizes the trajectory. Distributing this mass within a ring at the same distance does not have an effect over the first term in \eqref{eq: Omega curl}, which is a vector parallel to the symmetry axis. We would just have to keep the integral for longer.
\begin{align}
    \nonumber
	\vec{g}&=-4G\int d^3\vec{r'}\frac{m_1\delta^3(\vec{r'}-\vec{\tilde{r}}(\varphi))\vec{\tilde{v}}(\varphi)}{|\vec{r}-\vec{r'}|}\\
	&=-\frac{4Gm_1\,\vec{\tilde{v}}(\varphi)}{|\vec{r}-\vec{\tilde{r}}(\varphi)|}.
\end{align}

We now use Eq.~\eqref{eq: Omega curl}, first noting that
\begin{align}
    \nonumber
	\nabla\times\left(\frac{\vec{\tilde{v}}(\varphi)}{|\vec{r}-\vec{\tilde{r}}(\varphi)|}\right)&=-\vec{\tilde{v}}(\varphi)\times\nabla\cdot\left(\frac{1}{|\vec{r}-\vec{\tilde{r}}(\varphi)|}\right)\\
	&=\vec{\tilde{v}}(\varphi)\times\frac{\vec{r}-\vec{\tilde{r}}(\varphi)}{|\vec{r}-\vec{\tilde{r}}(\varphi)|^3},
\end{align}
where the first equality comes from the vectorial identity $\nabla\times(f\vec{a})=(\nabla f)\times\vec{a}+f(\nabla\times\vec{a})$, where in this case the second term is zero.

Taking this into account, we can use the Eq.~\eqref{eq: Omega curl} to get
\begin{align}
    \nonumber
	\vec{\Omega}&=2m_1\nabla\times\left(\frac{\vec{\tilde{v}}(\varphi)}{|\vec{r}-\vec{\tilde{r}}(\varphi)|}\right)\\
	&=2m_1\vec{\tilde{v}}(\varphi)\times\frac{\vec{r}-\vec{\tilde{r}}(\varphi)}{|\vec{r}-\vec{\tilde{r}}(\varphi)|^3}.
\end{align}

We can now get rid of the $\vec{r}$ by staying at the coordinate origin, $\vec{r}=0$, and thus get (dropping the tildes and the $\varphi$ dependence)
\begin{equation}
	\label{eq: Omega r v}
	\vec{\Omega}=2m_1\frac{\vec{r}\times\vec{v}}{r^3}.
\end{equation}

We will now make use of several equations from hyperbolic motion, namely
\begin{equation}
	r=a(e\cosh(E)-1),\qquad \vec{r}\times\vec{v}=a(e-1)v_0,
\end{equation}
where the second expression comes from angular momentum conservation and $E$ is the eccentric anomaly, related to the true anomaly by
\begin{equation}
    \tan^2\left(\frac{\varphi-\varphi_0}{2}\right)=\frac{e+1}{e-1}\tanh^2\left(\frac{E}{2}\right).
\end{equation}

As a result,
\begin{align}
    \nonumber
	\vec{\Omega}&=2m_1\frac{(e-1)v_0}{a^2(e\cosh(E)-1)^3}\\
	\label{eq: Omega hyperbolic}
	&=\frac{2m_1}{M^2}\frac{1}{(e\cosh(E)-1)^3}\frac{(e-1)^3}{(e+1)^2}v_0^5.
\end{align}

If we impose $E=0$, corresponding to the point of closest approach, $\varphi=\varphi_0$, and compute $\chi$, we get
\begin{equation}
	\chi=\frac{8}{5}m_2\frac{2m_1}{M^2}\frac{1}{(e+1)^2}v_0^5=\frac{16}{5}\frac{m_1m_2}{M^2}\frac{1}{(e+1)^2}v_0^5,
\end{equation}
which is, remarkably, the same expression we have obtained for the ring case, Eq.~\eqref{eq: chi ring}.

\subsection{From spin-orbit equations at PPN(1.5)}

Alternative to the previous approaches, we can use some equations from Ref.~\cite{De_Vittori_2014}. In this paper, the spins of two black holes in hyperbolic motion both follow precession dynamics given by the vectors
\begin{equation}
	\vec{\Omega_i}=\frac{\hat{k}}{M}\frac{\overline{\xi}^{5/3}\sqrt{e_t^2-1}}{(e_t\cosh(E)-1)^3}\delta_i,
\end{equation}
for $i=1,2$, where $\hat{k}$ is the unit vector perpendicular to the orbital angular momentum and
\begin{equation}
	\label{eq: deltas}
	\overline{\xi}=M\overline{n},\quad\delta_{1,2}=\frac{\eta}{2}+\frac{3}{4}(1\mp\sqrt{1-4\eta}),
\end{equation}
\begin{equation}
	\eta=m_1m_2/M^2,\quad m_1\geq m_2,
\end{equation}
with $\overline{n}$ being the mean motion of the hyperbolic orbit and $E$ its eccentric anomaly. Both $e_t$ and $\overline{n}$ are deviations of the keplerian case, taken at PPN(1.5) order. In our case, we will take them as if they were the exact newtonian values: $\overline{n}=n$, $e_t=e$. In standard hyperbolic motion, $n$ is given by the expression $n^2a^3=M$.

If we now substitute these expressions, we get
\begin{align}
    \nonumber
	\Omega_i&=\frac{1}{M}\frac{(M/a)^{5/2}\sqrt{e^2-1}}{(e\cosh(E)-1)^3}\delta_i\\
	\nonumber
	&=\frac{1}{M}\frac{\sqrt{e^2-1}}{(e\cosh(E)-1)^3}\frac{(e-1)^{5/2}}{(e+1)^{5/2}}\delta_iv_0^5\\
	&=\frac{1}{M}\frac{1}{(e\cosh(E)-1)^3}\frac{(e-1)^3}{(e+1)^2}\delta_iv_0^5,
\end{align}
which is somewhat similar to the expression~\eqref{eq: Omega hyperbolic}. The difference is a factor
\begin{equation}
	\frac{M}{2m_i}\delta_i.
\end{equation}

We can now compute the dimensionless spins, which in this case we have to split into two separate expressions. We also evaluate at $E=0$:
\begin{equation}
	\label{eq: chi PN 1}
	\chi_1=\left[\frac{4}{5}\eta+\frac{6}{5}\left(1-\sqrt{1-4\eta}\right)\right]\frac{m_1}{M}\frac{1}{(e+1)^2}v_0^5,
\end{equation}
\begin{equation}
	\label{eq: chi PN 2}
	\chi_2=\left[\frac{4}{5}\eta+\frac{6}{5}\left(1+\sqrt{1-4\eta}\right)\right]\frac{m_2}{M}\frac{1}{(e+1)^2}v_0^5.
\end{equation}

The comparison with the ring expression is difficult, due to factors dependent on the masses that were not present before. However, for the case $m_1=m_2$, both spins are equal and we get
\begin{equation}
	\label{eq: equal mass PN}
	\chi=\frac{7}{10}\frac{1}{(e+1)^2}v_0^5,
\end{equation}
which is formally identical to the ring case but with a factor 7/8 difference.

\subsection{Differences between the two expressions}

\begin{figure}[h]
\includegraphics[width=\linewidth]{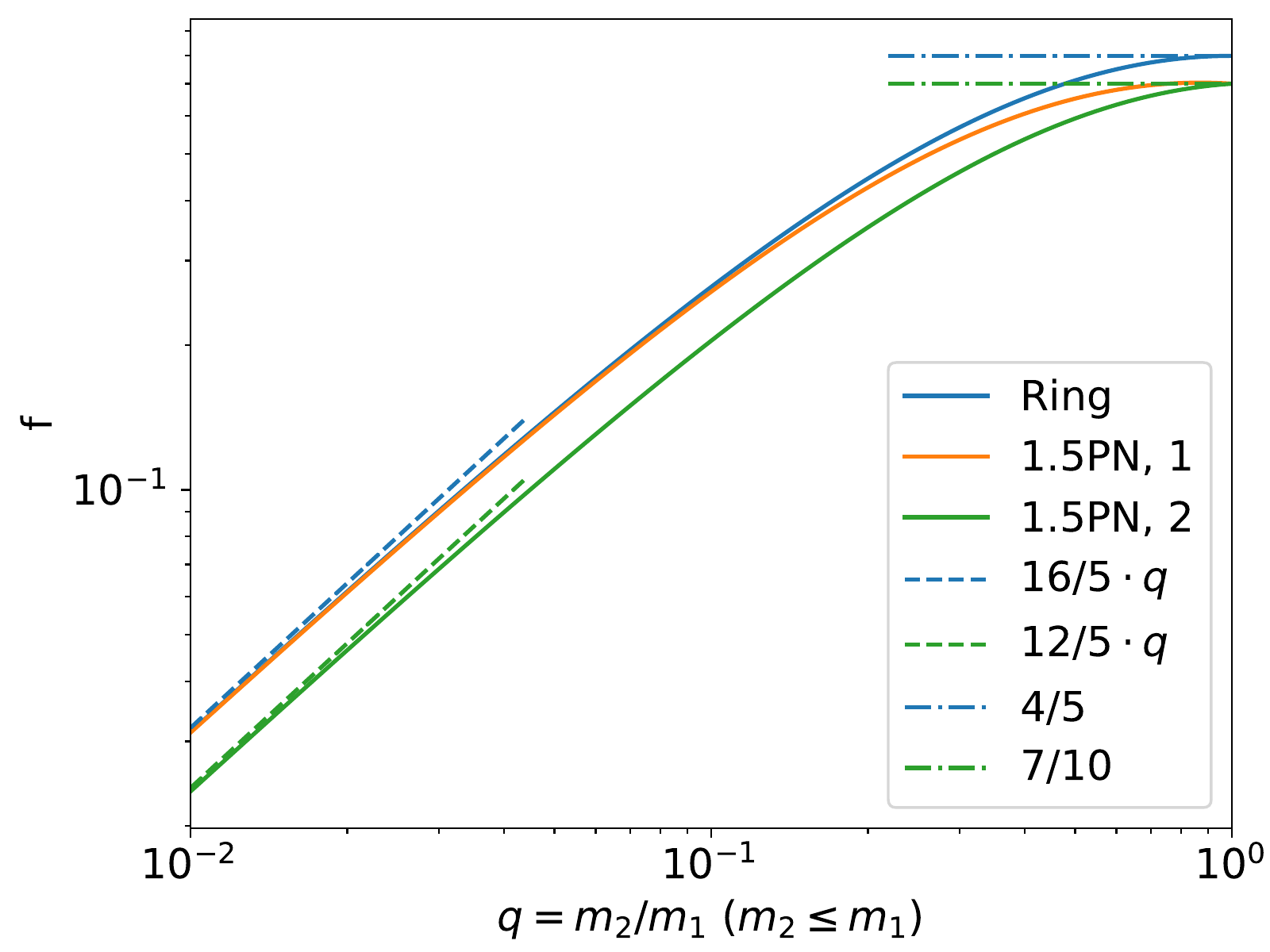}
\caption{\label{fig:fvsratio}Factor $f$ versus mass ratio $q=m_2/m_1$ ($m_2\leq m_1$). The trends for $q\to 0, 1$ are also provided for each curve.}
\end{figure}

In order to better understand the differences between the three expressions~\eqref{eq: chi ring},~\eqref{eq: chi PN 1} and~\eqref{eq: chi PN 2}, we can plot the factors $f$ that appear in these expressions before the $v_0^5/(e+1)^2$, which are given by Eqs.~\eqref{eq: chi ring},~\eqref{eq: chi PN 1} and~\eqref{eq: chi PN 2}. These values are shown in Fig.~\ref{fig:fvsratio}.

We have also plotted the asymptotic values to which each curve tends to for $q\to 0$. These can easily be found from the expressions of each factor in the limit $q\ll 1$, 
\begin{equation}
	f_{\rm ring},f_{\rm 1.5PN, 1}\to\frac{16}{5}q,\qquad f_{\rm 1.5PN,2}\to\frac{12}{5}q.
\end{equation}

\vspace*{2mm}

Interestingly, for the PPN(1.5) approximation, the induced spin on the most massive black hole approaches the same trend as that of the ring. Also, in the PPN(1.5) case, the highest spin is induced on the most massive black hole.

Fig.~\ref{fig:fvsratio} also shows that the order of magnitude of the computed spins is essentially the same, independent of the method that we use. In particular, the maximum difference between the different cases is a factor 4/3.

\bibliography{main}

\end{document}